\documentclass[12pt, prd, showpacs]{revtex4}
\usepackage{amssymb}
\usepackage{amsmath}

\input{tcilatex}

\begin{document}

\title{Collisions of massive and massless particles around rotating black
holes: general analysis}
\author{O. B. Zaslavskii}
\affiliation{Department of Physics and Technology, Kharkov V.N. Karazin National
University, 4 Svoboda Square, Kharkov, 61077, Ukraine}
\email{zaslav@ukr.net }

\begin{abstract}
We consider collisions between massive (electrons) and massless (photons)
particles near the horizon of a rotating black hole. Similarly to collisions
between massive particles, the infinite energy in the centre of mass frame
occurs in some situations. Namely, for one particle the relationship between
the energy and angular momentum should have a special form ("critical"
particle) whereas for the second one it should not hold ("usual" particle).
All combinations of possible pairs of critical and usual particles are
analyzed. The existence (or absence) of the effect is determined depending
on competition of two factors - gravitational blue shift for a photon
propagating towards a black hole and the Doppler effect due to
transformation from the locally nonrotating frame to a comoving one. Thus, a
pure kinematic explanation is suggested for the effect of infinitely growing
energies in the centre of mass frame.
\end{abstract}

\maketitle

\section{Introduction}

The effect of acceleration of particles by black holes up to arbitrary high
energies in the centre of mass frame was found recently \cite{ban} and
investigated further \cite{berti} - \cite{k}. In these works, mainly
collision between two massive particles was studied near the black hole
horizon. The case when one of particle is massless was considered in \cite%
{kerr} but for the concrete case of the Kerr metric only. The aim of the
present work is to give general analysis of acceleration of particles by
black holes due to near-horizon collisions between massive and massless
particles. In doing so, the main emphasis is made on a simple kinematic
approach. For massive particles it was developed in a previous work \cite{k}%
. Namely, if one particle (called critical) has a special relationship
between the energy and the angular momentum (or charge), its velocity in the
frame \ of ZAMO (zero angular momentum observer \cite{aj72}, also notation
LNRF is widely used - locally nonrotating frame) is less than the speed of
light even near the horizon. For a particle for which a similar relationship
does not hold, the velocity approaches that of light near the horizon. As a
result, it turns out that their relative velocity approaches that of light.
Therefore, an observer comoving with one particle, sees how the second one
moves with almost the speed of light and, thus, has unbound energy.

However, if one of particles is massless, explanation given in \cite{k} is
not valid since (i) there is no comoving frame for a massless particle, (ii)
their relative velocity is always equal to that of light. Therefore,
kinematic explanation should be somewhat changed. For brevity, we call
massive particle "electron" and massless one "photon", although
consideration applies to any kinds of such particles.

We do not consider the case when both particles are massless since we are
dealing with situation when both particles are ingoing and approach the
horizon. Then, collision between photons does not occur since in the regular
point two ingoing lightlike geodesics cannot intersect. (Collisions between
massive ingoing and outgoing particles near the horizon were considered in 
\cite{pir}.) Apart from this, a classical electrodynamics is linear theory,
so interaction between ingoing and outgoing photons could occur due to weak
quantum-electrodynamic effects only.

In what follows, we exploit the system of units which the speed of light $c=1
$. One reservation is in order. Hereafter, by the horizon limit we imply
only that the lapse function almost vanishes. Meanwhile, in the extremal
case the proper distance to the horizon itself remains, as is known,
infinite.

\section{Basic equations}

Let us consider the space-time of a rotating black hole described by the
metric

\begin{equation}
ds^{2}=-N^{2}dt^{2}+g_{\phi \phi }(d\phi -Cdt)^{2}+dl^{2}+g_{zz}dz^{2}.
\label{m}
\end{equation}%
We assume the existence of two Killing vectors, so the metric coefficients
do not depend on $t$ and $\phi $. On the horizon the lapse function $N=0$.
Instead of $l$ and $z$, one can use coordinates $\theta $ and $r$, which are
similar to the Boyer--Lindquist ones for the Kerr metric. In (\ref{m}) we
assume that the metric coefficients are even functions of $z$, so the
equatorial plane $\theta =\frac{\pi }{2}$ ($z=0$) is a symmetry one. We do
not specify the explicit form of the metric which is more general than the
Kerr one due to, say, matter that surrounds it.

We consider the geodesic motion of particles in the equatorial plane $\theta
=\frac{\pi }{2}$. For massive particles, the equations of motion read%
\begin{equation}
\dot{t}=u^{0}=\frac{E-\omega L_{1}}{N^{2}}\text{,}  \label{tm}
\end{equation}%
\begin{equation}
\dot{\phi}=\frac{L_{1}}{g_{\phi \phi }}+\frac{C(E-CL_{1})}{N^{2}},
\end{equation}%
\begin{equation}
\dot{l}^{2}=\frac{(E-CL_{1})^{2}}{N^{2}}-1-\frac{L_{1}^{2}}{g_{\phi \phi }}
\label{1m}
\end{equation}%
where $E=-u_{0}$ and $L_{1}=-u_{\phi }$ are conserved energy and angular
momentum per unit mass, $u^{\mu }$ is the four-velocity, dot denotes
differentiation with respect to the proper time.

For massless particles, the equations of motion have the form%
\begin{equation}
\frac{dt}{d\lambda }=k^{0}=\frac{\omega _{0}-CL_{2}}{N^{2}}\text{,}
\label{0}
\end{equation}%
\begin{equation}
\frac{d\phi }{d\lambda }=\frac{L_{2}}{g_{\phi \phi }}+\frac{C(\omega
_{0}-CL_{2})}{N^{2}},
\end{equation}%
\begin{equation}
\left( \frac{dl}{d\lambda }\right) ^{2}=\frac{(\omega _{0}-CL_{2})^{2}}{N^{2}%
}-\frac{L_{2}^{2}}{g_{\phi \phi }},  \label{1}
\end{equation}%
where $\omega _{0}=-k_{0}$, and $L_{2}=-k_{\phi }$ are conserved frequency
and angular momentum, $k^{\mu }$ is the wave vector, $\lambda $ is the
affine parameter. The quantity $\omega _{0}$ has a meaning of frequency
measured by a remote observer at infinity where we assume that $C\rightarrow
0$, $N\rightarrow 1$.

Thus, the only difference is in eqs. (\ref{1m}), (\ref{1}). We assume that $%
\dot{t}>0$, so that for the electron $E_{0}-CL>0$ (motion forward in time),
except, possibly on the horizon where we admit the equality $%
E_{0}-C_{H}L_{1}=0$ (subscript "H" denotes quantities calculated on the
horizon). For brevity, if $E-C_{H}L_{1}>0$ we call a particle "usual" and if 
$E-C_{H}L_{1}=0$ we call it "critical". In a similar way, a photon is
critical if $\omega _{0}-C_{H}L_{2}=0$.

\section{Energy in the centre of mass frame}

By definition, the energy $E_{c.m.}$ in the centre of mass frame is given by
the expression%
\begin{equation}
E_{c.m.}^{2}=-(p^{\mu }+k^{\mu })^{2}
\end{equation}%
where the Planck constant 
%TCIMACRO{\U{127}}%
%BeginExpansion
h{\hskip-.2em}\llap{\protect\rule[1.1ex]{.325em}{.1ex}}{\hskip.2em}%
%EndExpansion
=1, $p^{\mu }=mu^{\mu }$, $m$ is the electron rest mass. Then, 
\begin{equation}
E_{c.m.}^{2}=m^{2}-2m(uk)\text{, }(uk)\equiv u^{\mu }l_{\mu }\text{.}
\label{en}
\end{equation}%
It follows from (\ref{tm}) - (\ref{1}) that%
\begin{equation}
-(uk)=\frac{X}{N^{2}}-Y\text{,}  \label{uk}
\end{equation}%
where%
\begin{equation}
X=X_{1}X_{2}-Z_{1}Z_{2}\text{, }  \label{x}
\end{equation}%
$\,X_{1}\equiv E_{1}-CL_{1}$, $X_{2}=\omega _{0}-CL_{2}$,%
\begin{equation}
Z_{i}=\sqrt{X_{i}^{2}-N^{2}b_{i}}\text{, }b_{1}=1+\frac{L_{i}^{2}}{g_{\phi
\phi }}\text{, }b_{2}=\frac{L_{2}^{2}}{g_{\phi \phi }},  \label{z1}
\end{equation}%
\begin{equation}
Y=\frac{L_{1}L_{2}}{g_{\phi \phi }}\text{.}
\end{equation}

When, repeating the straightforward calculations along the lines of \cite%
{prd} step by step, one can arrive at the conclusions that unbound grow of. $%
E_{c.m.}^{2}$ is indeed possible if electron is critical, photon is usual or
vice versa.

For the extremal horizon one can obtain directly that in these situations $%
\lim_{N\rightarrow 0}E_{c.m.}^{2}=\infty $. For the nonextremal horizon, a
critical particle cannot reach it. However, if one takes $E-C_{H}L_{1}\sim
\delta \gtrsim N$, it turns out that $E_{c.m.}^{2}\sim \delta ^{-1}$ is
finite but can be made as large as one likes. This is completely similar to
the collisions of massive particles (see \cite{gp4}, \cite{prd}, \cite{gpm}
for details).

\section{Choice of frame}

Meanwhile, it is more important to obtain qualitative explanation of
infinite grow of $E_{c.m.}^{2}$ without explicit calculation of (\ref{uk}).
To this end, we introduce the tetrad basic that enables us to use the
formulas of special relativity locally, in the flat space-time tangent to a
given point. Denoting coordinates $x^{\mu }$ as \ $x^{0}=t,x^{1}=l$, $x^{2}=z
$, $x^{3}=\phi $, we choose the tetrad vectors $h_{(a)\mu }$ in the
following way (cf. \cite{aj72}):%
\begin{equation}
h_{(0)\mu }=-N(1,0,0,0)\text{, }  \label{h0}
\end{equation}%
\begin{equation}
h_{(1)\mu }=(0,1,0,0)
\end{equation}%
\begin{equation}
h_{(2)\mu }=\sqrt{g_{zz}}(0,0,0,1)
\end{equation}%
\begin{equation}
h_{(3)\mu }=\sqrt{g_{\phi \phi }}(-C,0,0,1)  \label{h3}
\end{equation}%
This is just the ZAMO frame. If such a tetrad is attached to an observer
moving in the metric (\ref{m}), it "rotates with the geometry" in the sense
that $\frac{d\phi }{dt}\equiv C$ for him.

Then, for an electron we can introduce the three-velocity according to \cite%
{aj72}%
\begin{equation}
v^{(i)}=v_{(i)}=\frac{u^{\mu }h_{\mu (i)}}{-u^{\mu }h_{\mu (0)}}.  \label{vi}
\end{equation}

One can check that

\begin{equation}
-u_{\mu }h_{(0)}^{\mu }=\frac{E-CL_{1}}{N},
\end{equation}%
\begin{equation}
u_{\mu }h_{(3)}^{\mu }=\frac{L_{1}}{\sqrt{g_{\phi \phi }}}\text{.}
\end{equation}%
\begin{equation}
v^{(1)}=\sqrt{1-\frac{N^{2}}{(E-CL_{1})^{2}}(1+\frac{L_{1}^{2}}{g_{\phi \phi
}})}  \label{v1}
\end{equation}%
\begin{equation}
v^{(3)}=\frac{L_{1}N}{\sqrt{g_{\phi \phi }}(E-CL)}  \label{v3}
\end{equation}

Then, after simple manipulations, one obtains \cite{k} that 
\begin{equation}
E-CL_{1}=mN\gamma \text{, }\gamma =\frac{1}{\sqrt{1-v^{2}}}\text{.}
\label{e}
\end{equation}%
where the absolute value of the velocity $v$ equals%
\begin{equation}
v^{2}=\left[ v^{(1)}\right] ^{2}+\left[ v^{(2)}\right] ^{2}.
\end{equation}

It is worth noting that formula (\ref{e}) can be inferred from eq. (88.9) of
the textbook \cite{ll} if one takes into account that $E_{1}-CL_{1}$ is just
the energy in the frame that rotates with the angular velocity $C$.

For an usual particle, in the horizon limit $N\rightarrow 0$, one has $%
v\rightarrow 1$, $E-C_{H}L_{1}\neq 0$. A critical particle can reach the
horizon in the extremal case only \cite{prd}, then taking into account the
main order $\omega -\omega _{H}\sim N$, one can see that in the horizon
limit, $E-CL_{1}\sim N$, $v\neq 1$.

In a similar way, we can obtain formulas for the photon. In contrast to (\ref%
{vi}), now formulas for the $k^{\mu }$ do not contain denominator:%
\begin{equation}
k^{(i)}=k_{(i)}=k^{\mu }h_{\mu (i)}\text{, }k^{(0)}=k^{\mu }h_{\mu
}^{(0)}=-k^{\mu }h_{\mu (0)}\text{.}
\end{equation}%
This is due to the fact that instead of the proper time $\tau $ the
parameter $\lambda $ along the geodesics is used, the vector $k^{\mu }$
being light-like.

From equations of motion (\ref{0}) - (\ref{1})\ and formulas for tetrad
components, we have 
\begin{equation}
k^{(1)}=-\sqrt{\omega ^{2}-\frac{L^{2}}{g_{\phi \phi }}}\text{,}  \label{k1}
\end{equation}

\begin{equation}
k^{(3)}=\frac{L}{\sqrt{g_{\phi \phi }}},  \label{k3}
\end{equation}%
where we took sign "-" in (\ref{k1}) since we consider an ingoing photon.
The analog of eq. (\ref{e}) reads%
\begin{equation}
\omega =\frac{\omega _{0}-CN}{N}\text{.}  \label{w}
\end{equation}%
It can be also obtained writing the scalar $(uk)$ in two frames - the
original system (\ref{m}) and the ZAMO one.

Defining $k^{2}=\left[ k^{(1)}\right] ^{2}+\left[ k^{(2)}\right] ^{2}$, it
is seen that \ 
\begin{equation}
k^{2}=\frac{(\omega _{0}-CL)^{2}}{N^{2}}=\omega ^{2}
\end{equation}%
\begin{equation}
k^{(0)}=-k_{\mu }h_{(0)}^{\mu }=\frac{\omega _{0}-LC}{N}=\omega
\end{equation}%
as it should be for the lightlike vector since $k^{2}-\left( k^{(0)}\right)
^{2}=0$.

In the horizon limit $N\rightarrow 0,$ the component $v^{(3)}\rightarrow 0$, 
$v^{(1)}\rightarrow 1$ for an usual electron. Therefore, the unit vector $%
\vec{n}_{1}=\frac{\vec{v}}{v}$ is pointed along $l$ direction,
perpendicularly to the horizon. For the critical particle this is not so 
\cite{ban} since $v^{(1)}\sim v^{(3)}$ have the same order. The similar
properties hold in the case of a photon for the vector $\vec{n}_{2}=\frac{%
\vec{k}}{k}$. Thus, in the horizon limit $(\vec{n}_{1}\vec{n}_{2})=1$ when
both particles are usual and $(\vec{n}_{1}\vec{n}_{2})\neq 1$ in other cases.

\section{Different types of collisions}

Now, we consider separately different cases depending on which particle (if
any) is critical.

\subsection{Case 1: electron is critical, photon is usual}

Let us pass to the frame which is comoving with respect to the electron.
Then, the frequency $\omega ^{\prime }$ measured in this frame is related to
the frequency $\omega $ in the ZAMO frame by the standard relativistic
formula%
\begin{equation}
\omega ^{\prime }=\gamma (\omega -\vec{k}\vec{v})=\omega \gamma \lbrack 1-v(%
\vec{n}_{1}\vec{n}_{2})]\text{. }  \label{om}
\end{equation}

For a critical particle, as is explained above and in \cite{k}, $v\neq 1$,
so the Lorentz factor $\gamma $ is finite. The scalar product $(\vec{n}_{1}%
\vec{n}_{2})\neq 1$, the quantity $\omega ^{\prime }$ has the order $\omega
. $ But, as a photon is usual, $\omega \rightarrow \infty $. Thus, $\omega
^{\prime }\rightarrow \infty $ as well, so the effect reveals itself.

The resulting effect can be interpreted as a consequence of two factors. On
one hand, there is an infinite blue shift of radiation due to strong
gravitating field near a black hole. From the other hand, there is red shift
due to the Doppler effect since a receiver of radiation is moving apart from
a photon. It turned out that in the case under discussion the first factor
is infinite whereas the second one is finite, so the net outcome is due to
blue shift.

\subsection{Case 2: electron is usual, photon is critical}

As the photon is critical,\thinspace\ $\omega $ is finite. But, as the
electron is usual, $v\rightarrow 1$, $\gamma \rightarrow \infty $. The
quantity $(\vec{n}_{1}\vec{n}_{2})\neq 1$. Thus, as a result, $\omega
^{\prime }\rightarrow \infty $ and we again obtain the effect under
discussion.

Interpretation again involves the Doppler effect but the concrete details
change. Let in a flat space-time a photon with the frequency $\omega $
propagate in the laboratory frame and some observer moves with the velocity $%
v$ with respect to this frame. Then, in in its own frame, the observer
measures the frequency of the process which is equal to $\omega ^{\prime }$.
In the case under discussion, $(\vec{n}_{1}\vec{n}_{2})\neq 1$. For
simplicity, we can take $(\vec{n}_{1}\vec{n}_{2})=0$. Then, the frequency
measured in the frame of a receiver $\omega ^{\prime }=\omega \gamma >\omega 
$ due to the transverse Doppler effect. In the limit $v\rightarrow 1$, the
Lorentz factor $\gamma \rightarrow 1$ and the frequency $\omega ^{\prime
}\rightarrow \infty $. In other words, even despite a moderate gravitational
blue shift that resulted in a finite $\omega $, the net outcome is infinite
due to the Doppler effect. 

\subsection{Case 3: both particles are critical}

Then, $(\vec{n}_{1}\vec{n}_{2})=1$ but $v<1,\omega $ is finite. It follows
from (\ref{om}) that $\omega ^{\prime }$ is also finite, so there is no
effect under discussion. In other words, both factors - gravitational blue
shifting and the Doppler effect are restricted and cannot give rises to
infinite energies.

\subsection{Case 4: both particles are usual}

Here, an accurate estimate of different terms in the horizon limit is
required. In the limit $N\rightarrow 0$ the quantities $\gamma \sim \frac{1}{%
N}$, $\omega \sim \frac{1}{N}$ as it is seen from (\ref{e}), (\ref{w}). It
follows also from (\ref{v1}), (\ref{v3}), (\ref{k1}), (\ref{k3}) that

\begin{equation}
1-(\vec{n}_{1}\vec{n}_{2})\sim N^{2}\text{.}
\end{equation}

As a result, the factors $N^{2}$ in the numerator and denominator compensate
each other, $\omega ^{\prime }$ remains finite, the effect of infinite
acceleration is absent. One can say that the effect of infinite red shift
due to Doppler effect for a receiver moving apart from the photon is
completely compensated by an infinite blue shifting the photon frequency.

\section{Conclusion}

Kinematic explanation given in the present article (see also its counterpart
for massive ones \cite{k}) showed the distinguished role of critical
particles: for electrons it means that their velocity in the ZAMO frame
remains less than the speed of light even near the horizon, for photons
their frequency in the same frame remains finite notwithstanding the
vanishing the lapse function near the horizon. The crucial point is that the
effect of infinite energies in the centre of mass frame is possible only for
the case when one (and only one) of colliding particles is critical. The
role of critical particles gave rise to natural classification taking into
account two factors - gravitational blue shift (GB) and the Doppler effect
(DE). Namely, we have four cases: 1) critical electron, usual photon:
infinite GB, finite DE, $E_{c.m.}$ is infinite, 2) critical photon, usual
electron: finite GB, infinite DE, $E_{c.m.}$ is infinite, 3) both particles
are critical: finite GB, finite DE, $E_{c.m.}$ is finite, 4) both particles
are usual: infinite GB, infinite DE, $E_{c.m.}$ is finite due to their
compensation.

The results of the present work can be used for investigation of the Compton
effect near black holes. Meanwhile, the possibility of infinite $E_{c.m.}$
means that, apart from mutual scattering of electrons and photons,
qualitatively new reactions can occur with creation of new kinds of high
energy particles.

In our consideration, the effects of backreaction and gravitational
radiation were neglected. There is physically attractive supposition \cite%
{berti}, \cite{ted} that these effects will change a picture drastically and
will restrict the grow of the energy $E_{c.m.}$. Meanwhile, the generality
of kinematic picture both for massive and massless particles altogether with
general geometric explanation \cite{cqg} means that this supposition should,
first of all, explain how the aforementioned effects could influence the
role of critical particles.


\begin{thebibliography}{99}
\bibitem{ban} M. Banados, J. Silk, S.M. West, Phys. Rev. Lett. \textbf{103},
111102 (2009).

\bibitem{berti} E. Berti, V. Cardoso, L. Gualtieri, F. Pretorius, U.
Sperhake, Phys. Rev.Lett. \textbf{103,} 239001 (2009).

\bibitem{ted} T. Jacobson, T.P. Sotiriou, Phys. Rev. Lett. \textbf{104},
021101 (2010).

\bibitem{lake1} K. Lake, Phys.\ Rev.\ Lett. \textbf{104}, 211102 (2010).

\bibitem{lake2} K. Lake, Phys.\ Rev.\ Lett. \textbf{104}, 259903(E) (2010).

\bibitem{gp4} A. A. Grib, Yu.V. Pavlov, Pis'ma v ZhETF, \textbf{92}, 147,
2010 (JETP Letters \textbf{92}, 125 (2010)).

\bibitem{kn} Shao-Wen Wei, Yu-Xiao Liu, Heng Guo,and Chun-E Fu, Phys.Rev.D 
\textbf{82},103005, (2010).

\bibitem{sen} Shao-Wen Wei, Yu-Xiao Liu, Hai-Tao Li, and Feng-Wei Chen, JHEP 
\textbf{12,} 066, 2010.

\bibitem{kk} Pu-Jian Mao, Ran Li, Lin-Yu Jia, Ji-Rong Ren, arXiv:1008.2660.

\bibitem{gpm} A. A. Grib and Yu.V. Pavlov, Gravitation Cosmol., \textbf{17},
42 (2011).

\bibitem{prd} O. B. Zaslavskii, Phys. Rev. D \textbf{82}, 083004 (2010).

\bibitem{jl} O. B. Zaslavskii, Pis'ma ZhETF \textbf{92}, 635 (2010) (JETP
Letters \textbf{92}, 571 (2010).

\bibitem{flux} M. Ba\~{n}ados, B. Hassanain, J. Silk and S. M. West \
Phys.Rev. D \textbf{83}, 023004 (2011).

\bibitem{bsw} Masashi Kimura, Ken-ichi Nakao and Hideyuki Tagoshi, Phys.Rev.
D \textbf{83}, 044013 (2011).

\bibitem{inter} Tomohiro Harada and Masashi Kimura, Phys. Rev. D \textbf{83}%
, 024002 (2011).

\bibitem{kerr} Tomohiro Harada and Masashi Kimura, Phys.Rev.D \textbf{83},
084041 (2011).

\bibitem{gs} Yi Zhu, Shao-Feng Wu, Yu-Xiao Liu, Ying Jiang, arXiv:1103.3848.

\bibitem{cqg} O. B. Zaslavskii, Classical Quantum Gravity, \textbf{28},
105010 (2011).

\bibitem{k} O. B. Zaslavskii, arXiv:1104.4802.

\bibitem{aj72} J. M. Bardeen, W.\ H. Press and S. A. Teukolsky, Astrophys.
Journ. \textbf{178}, 347 (1972).

\bibitem{pir} T. Piran and J. Shanam, Phys. Rev. D \textbf{16}, 1615 (1977).

\bibitem{lt} A. P. Lightman, W. H. Press, R. H. Price, and S. A. Teukolsky.
Problem book in Relativity and Gravitation. Princeton University Press.
Princeton. New Jerset, 1975.

\bibitem{ll} L. D. Landau and E.M. Lifshitz.\textit{Classical Theory of
Fields} (Butterworth-Heinemann; 1980).
\end{thebibliography}
\end{document}